\documentclass[aps,prb,twocolumn,showpacs]{revtex4}
\usepackage{graphicx}
\begin{document}
\title{Magnetoresistance oscillations in two-dimensional electron systems under \\ 
monochromatic and bichromatic radiations}
\author{X. L. Lei and S. Y. Liu}
\affiliation{Department of Physics, Shanghai Jiaotong University,
1954 Huashan Road, Shanghai 200030, China}

\begin{abstract}

The magnetoresistance oscillations in high-mobility 
two-dimensional electron systems induced by two radiation fields 
of frequencies 31\,GHz and 47\,GHz, are analyzed 
in a wide magnetic-field range down to 100\,G, using 
the balance-equation approach to magnetotransport for high-carrier-density systems. 
The frequency mixing processes are shown to be important. 
The predicted peak positions, relative heights, radiation-intensity dependence and 
their relation with monochromatic resistivities  
are in good agreement with recent experimental finding 
[M. A. Zudov {\it et al.}, Phys. Rev. Lett. {\bf 96}, 236804 (2006)].

\end{abstract}


\maketitle

The fascinating phenomena of radiation-induced magnetoresistance oscillations (RIMOs) 
and zero-resistance states (ZRS) in high mobility two-dimensional (2D) electron 
systems,\cite{Ryz,Zud01,Ye,Mani,Zud03} continue to attract much experimental\cite{Yang,Dor03,
Mani04,Will,Mani-apl,Zud04,Stud,Dor05,Mani05,Smet05,Zud06,Zud06prl} 
and theoretical\cite{Andreev,Durst,Xie,Lei03,Ryz05199,
Vav,Mikh,Dietel,Torres,Dmitriev04,Inar,Lei05,Ryz05,Joas05,Ng,Lei06prb,Kun08607} attention.

Under the influence of a monochromatic microwave radiation of frequency $\omega/2\pi$,
the low-temperature magnetoresistance of a 2D electron gas (EG), 
exhibits periodic oscillation as a function of the inverse magnetic field $1/B$.
The oscillation features the periodical appearance of peak-valley pairs 
around node points $\omega/\omega_c=j$ ($j=1,2,3,4,\cdots$, $\omega_c=eB/m$ 
is the cyclotron frequency). With increasing the radiation intensity 
the oscillation amplitude increases and the measured resistance around the minima of a few lowest-$j$ 
pairs can drop down to zero. The nature of 
this observed ZRS, which was considered as the result of absolute negative 
resistance,\cite{Andreev} remains an issue for experimental confirmation. 
A recent measurement of photoresistance under simultaneous illumination
of two radiation fields of different frequencies by Zudov {\it et al.}\cite{Zud06prl}
provided a plausible evidence to link the ZRS to the negative resistance. 
However, the empirical superposition relation between the bichromatic and
two monochromatic resistances, needs refinement and justification.\cite{Zud06prl} 
The present letter reports such an examination with the balance-equation approach\cite{Lei85}
to magnetotransport for high-carrier-density systems.

In this photon-assisted-scattering based theory,\cite{Lei03,Lei05} 
the transport state of a 2DEG subject to a dc field ${\bf E}_0$ and a
normally incident monochromatic radiation  
${\bf E}_{\rm i}={\bf E}_{{\rm i}s}
\sin(\omega t)+ {\bf E}_{{\rm i}c}\cos(\omega t)$, 
is described by the electron drift velocity
$
{\bf v_0}+ {\bf v}_{c} \cos(\omega t)+
{\bf v}_{s} \sin(\omega t) 
$
and an electron temperature $T_{\rm e}$. 
They are determined by the balance equations
(8)--(11) in Ref.\,\onlinecite{Lei05}. 
The linear (${\bf v}_0\rightarrow 0$) magnetoresistivity is expressed as 
\begin{equation}
R_{xx}^{\omega}(\xi)=-\frac{1}{N_{\rm e}^2 e^2}\sum_{{\bf q}_\|}q_x^2 |
U({\bf q}_\|)| ^2\sum_{n=-\infty }^\infty {J}_n^2(\xi) 
\,\Pi_2^{\prime}({\bf q}_\|, n\omega), \label{rxxm}
\end{equation}
where ${\bf q}_\|\equiv(q_x,q_y)$, $N_{\rm e}$ is the electron density, 
$U({\bf q}_\|)$ is the impurity potential,
$
\xi\equiv \sqrt{({\bf q}_\|\cdot {\bf v}_{c})^2+
({\bf q}_\|\cdot {\bf v}_{s})^2}/{\omega},
$ 
and $\Pi_2^{\prime}({\bf q}_\|, \Omega)\equiv 
\frac {\partial} {\partial\, \Omega }\Pi_2({\bf q}_\|,\Omega)$ stands for 
the differentiation of the imaginary part of the electron density correlation function 
with respect to its frequency variable.

The $\Pi_2({\bf q}_{\|}, \Omega)$ function of a 2D
system in a magnetic field can be expressed 
as sums over all Landau levels:
\begin{eqnarray}
&&\hspace{-0.7cm}\Pi _2({\bf q}_{\|},\Omega ) =  \frac 1{2\pi
l_{\rm B}^2}\sum_{l,l'}C_{l,l'}(l_{\rm B}^2q_{\|}^2/2) 
\Pi _2(l,l',\Omega),
\label{pi_2}\\
&&\hspace{-0.7cm}\Pi _2(l,l',\Omega)=-\frac2\pi \int d\varepsilon
\left [ f(\varepsilon )- f(\varepsilon +\Omega)\right ]\nonumber\\
&&\,\hspace{2cm}\times\,\,{\rm Im}G_l(\varepsilon +\Omega){\rm Im}G_{l'}(\varepsilon ), \label{pi_2ll}
\end{eqnarray}
where
$
C_{l,l+m}(Y)\equiv l![(l+m)!]^{-1}Y^m e^{-Y}[L_l^m(Y)]^2
$
with $L_l^m(Y)$ the associate Laguerre polynomial, 
$l_{\rm B}\equiv\sqrt{1/|eB|}$ is the magnetic length, 
$f(\varepsilon )=\{\exp [(\varepsilon -\mu)/T_{\rm e}]+1\}^{-1}$ 
is the Fermi distribution function at electron temperature $T_{\rm e}$. 
The density-of-states (DOS) of the $l$-th Landau level
is modeled with a Gaussian form,
${\rm Im}G_l(\varepsilon)=-(\sqrt{2\pi}/\Gamma)
\exp[-{2(\varepsilon-\varepsilon_l)^2}/{\Gamma^2}], 
$
having a half-width
$
\Gamma=\left(8e\omega_c\alpha/\pi m \mu_0\right)^{1/2}
$
around the level center $\varepsilon_l=l\omega_c$. Here $m$ is the carrier 
effective mass, 
$\mu_0$ is the linear mobility at lattice temperature $T$ 
in the absence of the magnetic field, 
and $\alpha$ is a semiempirical 
parameter.

It can be seen from expressions (\ref{pi_2}) and (\ref{pi_2ll}) 
that, in the case of low electron temperature ($T_{\rm e}\ll \epsilon_{\rm F}$, 
the Fermi level) 
and large Landau-level filling factor $\nu$, for
any integer $N$ satisfying $|N|\ll\nu$,
$\Pi_2^{\prime}({\bf q}_\|, \Omega+N\omega_c)=\Pi_2^{\prime}({\bf q}_\|, \Omega)$. Thus, 
 when $\omega=l\omega_c$ ($l=1,2,....$), the 
$\Pi_2^{\prime}({\bf q}_\|, n\omega)$ function in (\ref{rxxm}) is  
independent of $n$, that
$
\sum_{n}{J}_n^2(\xi)\, \Pi_2^{\prime}({\bf q}_\|, n\omega) 
= \Pi_2^{\prime}({\bf q}_\|, 0) 
$,
and
\begin{equation}
R_{xx}^{\omega}(\xi)=-\frac{1}{N_{\rm e}^2 e^2}\sum_{{\bf q}_\|}q_x^2 |
U({\bf q}_\|)| ^2 
\,\Pi_2^{\prime}({\bf q}_\|, 0). \label{rxxcyl} 
\end{equation}
(\ref{rxxcyl}) indicates
that at cyclotron frequency and its harmonics, $\omega=l\omega_c\, (l=1,2,...)$,
the difference of $R_{xx}^{\omega}(\xi)$ driven by monochromatic frequency $\omega$ fields 
of different strengths, can
appear only through the difference of electron heating contained in the $\Pi_2$ function.\cite{Lei03apl} 
Under the condition of $T_{\rm e}\ll\epsilon_{\rm F}$, 
the effect of electron heating on photoresistance 
is quite weak,\cite{Lei03} therefore, at  
integer-$\omega_c$ value of frequency $\omega$, the magnetoresistivity with radiation 
is essentially the same as that without radiation, $R_{xx}^{\omega}(\xi)\simeq R_{xx}(0)$, 
or the photoresponse vanishes at cyclotron resonance and its harmonics.

\begin{figure}
\includegraphics [width=0.45\textwidth,clip=on] {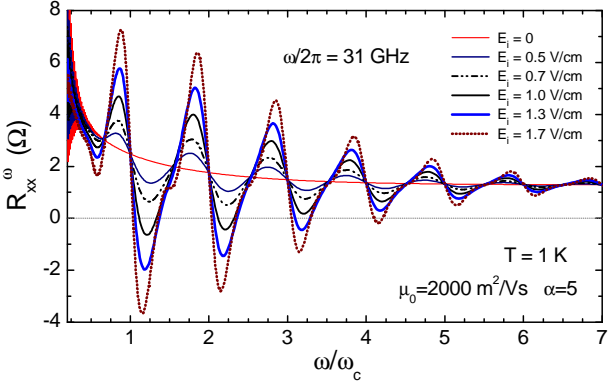}
\vspace*{-0.3cm}
\caption{(Color online) The magnetoresistivity $R_{xx}^{\omega}$
induced by monochromatic radiations of frequency $\omega/2\pi=31$\,GHz
at lattice temperature $T=1$\,K, 
in a GaAs-based 2DEG with $N_{\rm e}=2.4\times 10^{15}$\,m$^{-2}$, 
$\mu_0=2000$\,m$^2$/Vs and $\alpha=5$.}
\label{fig1}
\end{figure}

This is clearly seen in Fig.\,1, where we plot the calculated 
magnetoresistivity $R_{xx}^{\omega}$ of a GaAs-based 2DEG with 
$N_{\rm e}=2.4\times 10^{15}$\,m$^{-2}$, $\mu_0=2000$\,m$^2$/Vs and $\alpha=5$, 
irradiated by linearly polarized monochromatic waves of frequency $\omega/2\pi=31$\,GHz
 having incident amplitudes 
$E_{{\rm i}}=0,0.5,0.7,1.0,1.3$ and 1.7\,V/cm at lattice temperature $T=1$\,K.
The scattering is assumed due to short-range impurities and the material parameters used are the same 
as in Ref.\,\onlinecite{Lei06prb}. All the curves, with and without radiation,
cross at $\omega/\omega_c=1,2,3,4,5,6$ and 7.
This feature, being pointed out before 
theoretically,\cite{Durst,Xie,Lei03,Ryz05199} 
has been confirmed by experiments.\cite{Mani,Zud03,Mani04,Mani-apl}
It is quite general, independent of the polarization state
of the radiation, the behavior of the elastic scattering, and other properties of
2DEG, and holds quite precisely up to rather strong radiation field 
even multiphoton processes play important role, as long as the  
the 2DEG remains in degenerate ($T_{\rm e}\ll \epsilon_{\rm F}$). 
It thus provides a convenient and accurate method to determine 
the effective mass of 2DEG.

Another feature which shows up when the radiation strength is modest 
that contributions from two and higher photon processes are negligible and RIMOs result 
mainly from single-photon-assisted processes:
\begin{eqnarray}
&&\hspace*{-0.8cm}R_{xx}^{\omega}(\xi)\simeq -\frac{1}{N_{\rm e}^2 e^2}\sum_{{\bf q}_\|}q_x^2 |
U({\bf q}_\|)| ^2  
\left[{J}_0^2(\xi)\,\Pi_2^{\prime}({\bf q}_\|, 0)\right.\nonumber\\
&&\hspace{0.8cm}\left.+{J}_1^2(\xi)\,\Pi_2^{\prime}({\bf q}_\|, \omega)
+{J}_{-1}^2(\xi)\,\Pi_2^{\prime}({\bf q}_\|,-\omega)\right].
\end{eqnarray}
Since the DOS function decays rapidly when deviating 
from the center of each Landau level, the magnitude of 
$\Pi_2^{\prime}({\bf q}_\|, \omega)$ reaches a minimum of almost
zero when $\omega/\omega_c=(l+\frac{1}{2})\,\,(l=1,2,3...)$. Therefore around 
these half-integer-$\omega_c$ positions, the magnetoresistivity 
\begin{equation}
R_{xx}^{\omega}(\xi)\simeq -\frac{1}{N_{\rm e}^2 e^2}\sum_{{\bf q}_\|}q_x^2 |
U({\bf q}_\|)| ^2 {J}_0^2(\xi)\,\Pi_2^{\prime}({\bf q}_\|, 0)
\end{equation}
is only weakly dependent on radiation strength through ${J}_0^2(\xi)$,
i.e. half-integer-$\omega_c$ positions are approximately node points for modest radiation.  
This can also been seen in Fig.\,1 where almost all the resistivity curves cross
at $\omega/\omega_c=3.5,4.5, 5.5$ and 6.5, and  
the curves of lower radiation strengths cross at $\omega/\omega_c=1.5$ and 2.5. 

Under normally illuminating bichromatic radiation 
${\bf E}_{\rm i1}={\bf E}_{{\rm i1}s}\sin(\omega_1 t)+ 
{\bf E}_{{\rm i1}c}\cos(\omega_1 t)$ 
and  
${\bf E}_{\rm i2}={\bf E}_{{\rm i2}s}\sin(\omega_2 t)+ 
{\bf E}_{{\rm i2}c}\cos(\omega_2 t)$,
the transport state of a 2DEG can be described by the electron drift velocity 
oscillating at both base radiation frequencies,
${\bf v}_0+{\bf v}_{1c} \cos(\omega_1 t)+{\bf v}_{1s} \sin(\omega_1 t)+
{\bf v}_{2c} \cos(\omega_2 t)+{\bf v}_{2s} \sin(\omega_2 t)$,
together with the electron temperature $T_{\rm e}$. 
They are determined by the balance equations
(5)--(10) in Ref.\,\onlinecite{Lei06prb}. 
The linear magnetoresistivity is expressed as 
\begin{eqnarray}
&&\hspace*{-0.5cm}R_{xx}^{\omega_1\omega_2}(\xi_1,\xi_2)=
-\frac{1}{N_{\rm e}^2 e^2}\sum_{{\bf q}_\|} q_x^2|
U({\bf q}_\|)|^2 \nonumber\\
&&\times\sum_{n_1,n_2=-\infty }^\infty  
J_{n_1}^2(\xi_1)J_{n_2}^2(\xi_2)
\, \Pi_2^{\prime}({\bf q}_\|, n_1\omega_1+n_2\omega_2).\hspace*{0.4cm}
\label{rxxb}
\end{eqnarray}
where 
$
\xi_1\equiv \sqrt{({\bf q}_\|\cdot {\bf v}_{1c})^2+
({\bf q}_\|\cdot {\bf v}_{1s})^2}/{\omega_1}
$ and 
$
\xi_2\equiv \sqrt{({\bf q}_\|\cdot {\bf v}_{2c})^2+
({\bf q}_\|\cdot {\bf v}_{2s})^2}/{\omega_2}.
$
(\ref{rxxb}) indicates that bichromatic $R_{xx}^{\omega_1\omega_2}$ 
contains, besides the weighted superposition of mono-$\omega_1$ and
mono-$\omega_2$ contributions,
$J_0^2(\xi_1) R_{xx}^{\omega_2}(\xi_2)+J_0^2(\xi_2) R_{xx}^{\omega_1}(\xi_1)$,
also mixing $\omega_1$ and $\omega_2$ contributions (term with $n_1=n_2=0$ and terms
with both $n_1, n_2\neq 0$).

In the degenerate and large-$\nu$ case,
at positions $\omega_2=l \omega_c \,\,(l=1,2,3....)$,
$\Pi_2^{\prime}({\bf q}_\|, n_1\omega_1+n_2\omega_2)$ function 
does not depend on $n_2$, thus (\ref{rxxb}) reduces to
\begin{eqnarray}
&&\hspace*{-0.5cm}R_{xx}^{\omega_1\omega_2}(\xi_1,\xi_2)=
-\frac{1}{N_{\rm e}^2 e^2}\sum_{{\bf q}_\|} q_x^2|
U({\bf q}_\|)|^2 \nonumber\\
&&\times\sum_{n_1=-\infty }^\infty  
J_{n_1}^2(\xi_1)
\, \Pi_2^{\prime}({\bf q}_\|, n_1\omega_1)\approx R_{xx}^{\omega_1}(\xi_1),\hspace*{0.4cm}
\label{rxxb2}
\end{eqnarray}
i.e. bichromatic $R_{xx}^{\omega_1\omega_2}(\xi_1,\xi_2)$ equals the monochromatic $R_{xx}^{\omega_1}(\xi_1)$
under $\omega_1$-${\bf E}_{\rm i1}$ radiation  
(effect of electron temperature difference on photoresistance is negligible).
This means that all the bichromatic $R_{xx}^{\omega_1\omega_2}(\xi_1,\xi_2)$ curves 
with fixed ${\bf E}_{\rm i1}$ but changing ${\bf E}_{\rm i2}$, 
cross with the monochromatic $R_{xx}^{\omega_1}(\xi_1)$ curve 
of ${\bf E}_{\rm i1}$ at integer-$\omega_c$ points of $\omega_2$.
Likewise, at integer-$\omega_c$ points of $\omega_1$, 
$R_{xx}^{\omega_1\omega_2}(\xi_1,\xi_2)
\approx R_{xx}^{\omega_2}(\xi_2)$.

\begin{figure}
\includegraphics [width=0.45\textwidth,clip=on] {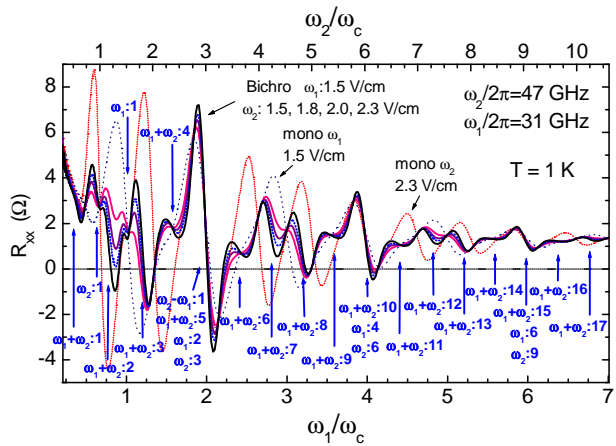}
\vspace*{-0.3cm}
\caption{(Color online) The magnetoresistivity $R_{xx}$ induced by 
bichromatic or monochromatic radiations ($R_{xx}^{\omega_1\omega_2}$,
$R_{xx}^{\omega_1}$ or $R_{xx}^{\omega_2}$) in the same 2DEG 
as described in Fig.\,1.}
\label{fig2}
\end{figure}
\begin{figure}
\includegraphics [width=0.45\textwidth,clip=on] {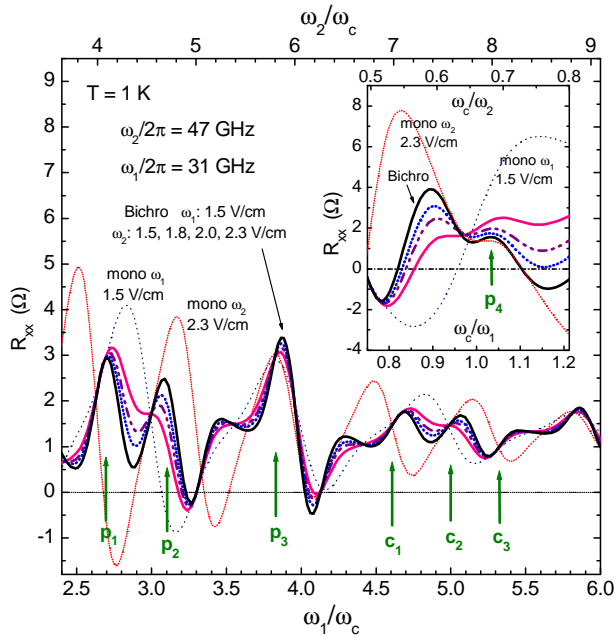}
\vspace*{-0.3cm}
\caption{(Color online) Same as figure 2 in an enlarged scale.}
\label{fig3}
\end{figure}

In the case of $\omega_2 \sim 1.5\omega_1$, even and odd $\omega_c$ points exhibit
somewhat different behavior.
At an even-$\omega_c$ point of $\omega_1$ 
($\omega_1=2\omega_c$, $4\omega_c$ or $6\omega_c$),  
$\omega_2$ is also close to an integer $\omega_c$ 
($\omega_2\approx 3\omega_c$, $6\omega_c$ and $9\omega_c$) that 
$R_{xx}^{\omega_2}(\xi_2)$ is almost identical to the dark resistivity $R_{xx}(0)$,
independent of ${\bf E}_{\rm i2}$. Therefore, 
all bichro-resistivity $R_{xx}^{\omega_1\omega_2}$, 
mono-resistivity $R_{xx}^{\omega_1}$ and mono-resistivity
$R_{xx}^{\omega_2}$,
cross at these points with dark resistivity $R_{xx}(0)$.
At an odd-$\omega_c$ point of $\omega_1$ 
($\omega_1=\omega_c$, $3\omega_c$, $5\omega_c$ or $7\omega_c$),  
$\omega_2$ is close to a half-integer $\omega_c$ 
($\omega_2\approx 1.5\omega_c$, $4.5\omega_c$, $7.5\omega_c$ or $10.5\omega_c$).
It is a node point of $R_{xx}^{\omega_2}(\xi_2)$ when 
$E_{\rm i2}$ changes modestly. Thus, at these points 
bichromatic $R_{xx}^{\omega_1\omega_2}$ and 
monochromatic $R_{xx}^{\omega_1}$ and 
 $R_{xx}^{\omega_2}$ are nearly the same,
but may be somewhat different from the dark resistivity. 

Figure 2 demonstrates the calculated magnetoresistivity $R_{xx}^{\omega_1\omega_2}$ 
versus the inverse magnetic field for the same GaAs-based 2D system as described 
in Fig.\,1 under linearly polarized bichromatic radiation of $\omega_1/2\pi=31$\,GHz
and $\omega_2/2\pi=47$\,GHz with incident amplitudes $E_{\rm i1}=1.5$\,V/cm and  
$E_{\rm i2}=1.5,1.8,2.0,$ and 2.3\,V/cm at $T=1$\,K.
The monochromatic $R_{xx}^{\omega_1}$ at $E_{\rm i1}=1.5$\,V/cm and
$R_{xx}^{\omega_2}$ at $E_{\rm i2}=2.3$\,V/cm are also shown. 
Comparing with monochromatic ones, the bichromatic curve consists of 
much more sizable peak-valley pairs. Since only minor high-order structures
showing up in both monochromatic curves at these radiation strengths, 
most of the pairs appearing in the bichromatic curves
relate to mixing photon processes.
We have identified the lowest order ($|n_1|\leq 1$ and $|n_2|\leq 1$)  
individual ($\omega_1$ or $\omega_2$) and mixing 
($\omega_1-\omega_2$ or $\omega_1+\omega_2$) 
photon-assisted electron transition processes 
to all the pairs in the figure. Except those pairs around $\omega_2/\omega_c=1$ 
and $\omega_1/\omega_c=1$, which relate to single-$\omega_2$ and single-$\omega_1$ 
processes, and those around $\omega_1/\omega_c=2, 4$ and 6, 
which relate to both individual and mixing photon processes, all other pairs are related 
to mixing processes $\omega_1+\omega_2:l$ ($l=1,2,...17$) with electron transitions jumping $l$
Landau-level spacings.\cite{Lei06prb}

A lower magnetic-field portion of this figure is redrawn in Fig.\,3 
with enlarged scale for clearer comparison with experiments. 
The predicted positions and relative heights
of bichromatic peaks $p_1$, $p_2$ and $p_3$ around $\omega_1/\omega_c=2.7,3.1$ and 3.9, and the predicted 
approximate crossing points $c_1$ (at $\omega_2/\omega_c=7$), $c_2$ 
(at $\omega_1/\omega_c=5$) and $c_3$ (at $\omega_2/\omega_c=8$),
well reproduce\cite{note} the experimental bichromatic peaks around
270, 222 and 182\,G and the experimental crossing points marked in Fig.\,2
of Ref.\,\onlinecite{Zud06prl}, where the empirical superposition relation
is considered valid. The portion of $\omega_c/\omega_1=1$ vicinity
of Fig.\,2 is also replotted in the inset of Fig.\,3 as functions of $B$. 
A bichromatic peak $p_4$ shows up clearly around $\omega_c/\omega_1=1.03$,
where mono-$\omega_1$ curve has a large positive value and mono-$\omega_2$ curve
exhibits a small positive peak. This is exactly what observed 
experimentally around $B\approx 730$\,G, where the superposition
relation is apparently invalid.\cite{Zud06prl}
The present theory, instead, is able to provide a unified description 
for the bichromatic resistivity and its relation to individual monochromatic ones 
over the wide magnetic field range.

This work was supported by Projects of the National Science Foundation of China
and the Shanghai Municipal Commission of Science and Technology.


\begin{thebibliography}{99}
\bibitem{note} If $\omega_c/\omega_1=1.03$ is considered at $730$\,G, 
the $p_1$, $p_2$ and $p_3$ locate respectively at 263, 229 and 182\,G.
\end{thebibliography}
\end{document}